\documentclass[aps,twocolumn,prb,showpacs,reprint,amsmath,amssymb,superscriptaddress]{revtex4}

\usepackage{graphicx}
\usepackage{dcolumn}
\usepackage{bm}

\begin{document}

\title{Origin of the hysteresis in bilayer 2D systems in the quantum Hall regime}

\author{L.H. Ho}
\email{laphang@phys.unsw.edu.au}
\affiliation{School of Physics, University of New South Wales,
Sydney NSW 2052, Australia}
\affiliation{CSIRO Materials Science and Engineering, P.O. Box 218, Lindfield NSW 2070, Australia}

\author{L.J. Taskinen}
\affiliation{School of Physics, University of New South Wales, Sydney NSW 2052, Australia}

\author{A.P. Micolich}
\email{adam.micolich@gmail.com} \affiliation{School of Physics,
University of New South Wales, Sydney NSW 2052, Australia}

\author{A.R. Hamilton}
\affiliation{School of Physics, University of New South Wales,
Sydney NSW 2052, Australia}

\author{P. Atkinson}
\affiliation{Cavendish Laboratory, University of Cambridge,
Cambridge CB3 0HE, United Kingdom}

\author{D.A. Ritchie}
\affiliation{Cavendish Laboratory, University of Cambridge,
Cambridge CB3 0HE, United Kingdom}

\date{\today}

\begin{abstract}

The hysteresis observed in the magnetoresistance of bilayer 2D
systems in the quantum Hall regime is generally attributed to the
long time constant for charge transfer between the 2D systems due to
the very low conductivity of the quantum Hall bulk states. We report
electrometry measurements of a bilayer 2D system that demonstrate
that the hysteresis is instead due to non-equilibrium induced
current. This finding is consistent with magnetometry and electrometry measurements
of single 2D systems, and has important ramifications for
understanding hysteresis in bilayer 2D systems.

\end{abstract}


\maketitle

The study of two-dimensional electron systems (2DESs) has led to
much interesting physics including the Nobel prize winning
discoveries of the integer~\cite{vonKlitzingPRL80} and
fractional~\cite{TsuiPRL82} quantum Hall effects. These effects are
dramatic examples of the vital role that dimensionality,
quantization and electron-electron interactions play in the physics
of semiconductor devices. Deeper insight into the role of
interactions can be obtained by locating a second 2DES in close
proximity. These `bilayer' 2D electron systems have received
significant interest as routes to realizing exotic new electronic
ground states ranging from excitonic Bose-Einstein
condensates~\cite{EisensteinNat04} through to quantum Hall
ferromagnets.~\cite{GirvinPhysTod00}

The latter has been a particular focus for attention with numerous
reports of hysteresis -- a common hallmark of ferromagnetic behavior
-- in bilayer 2D systems in the quantum Hall
regime.~\cite{PiazzaNat99, dePoortereSci00, ZhuPRB00, TutucPRB03,
PanPRB05, MisraPRB08} Bilayer 2D systems provide an additional layer
of complexity to studies of magnetic ordering because in addition to
the electron spin, these systems have a `pseudospin' degree of
freedom corresponding to which of the two layers an electron
occupies. For example, Piazza {\it et al.} studied a wide quantum
well containing two 2DESs, and showed that a first-order magnetic
phase transition could be induced by careful tuning of the energetic
alignment between a spin-up Landau level in one 2DES and a spin-down
Landau level in the other, an effect requiring both spin and
pseudospin to explain.~\cite{PiazzaNat99}

Subsequent papers have reported more widespread hysteresis. Zhu {\it
et al.} measured the longitudinal magnetoresistance $R_{xx}$ of a
2DES with a nearby, parallel impurity channel and observed
resistance spikes coinciding with a number of the integer quantum
Hall minima, and hysteresis between data obtained with
increasing/decreasing magnetic field at the edges of these
minima.~\cite{ZhuPRB00} Similar behavior was reported by Tutuc {\it
et al.}~\cite{TutucPRB03} and Misra {\it et al.}~\cite{MisraPRB08}
in bilayer 2D hole systems and by Pan {\it et al.}~\cite{PanPRB05}
in a bilayer 2DES. In each case, the authors explained these effects
as being entirely due to impeded charge transfer between the two 2D
systems, which occurs in the following way. At integer filling
factor, each 2DES will undergo a sudden change in its chemical
potential due to Landau level population/depopulation. This leads to
a non-equilibrium imbalance in the chemical potentials of the two 2D
systems, which is overcome by charge migration via the ohmic
contacts connecting the two 2D systems. However, this charge
migration is impeded by the localization of the bulk states in the
quantum Hall regime, leading to a large $RC$ time-constant, and
hence the observed hysteresis.~\cite{ZhuPRB00}

The problem with this {\it charge transfer} mechanism is that it
ignores another well established cause of hysteresis in 2D systems
in the quantum Hall regime -- {\it non-equilibrium induced currents}. 
%
Large and long-lived induced currents can circulate within the dissipationless edge states associated with the quantum Hall effect. These currents can be driven for instance by the induced emf caused by a changing magnetic field. 
%
They produce a hysteretic magnetization signal that has been widely observed in magnetometry studies of single layer 2D systems,~\cite{UsherJPCM09} both
in response to a changing magnetic field $B$,~\cite{WattsPRL98} and
changing density in the 2D system at constant
$B$.~\cite{FaulhaberPRB05} These induced currents also generate a large hysteretic electrostatic potential (of the order $\sim 10$~mV) that has been observed in single-electron transistor electrometry studies of single 2D
systems.~\cite{HuelsPRB04}
Thus it would be
surprising if non-equilibrium induced current did not play some
role, perhaps even a dominant one, in the hysteresis reported in
bilayer 2D systems.~\cite{ZhuPRB00,TutucPRB03,PanPRB05,MisraPRB08}

It is perhaps understandable that non-equilibrium induced current in
bilayer 2D systems has received limited atttention to this point, because
preceding studies have relied solely on traditional transport
measurements, which cannot provide any direct experimental evidence
for the relative contributions of the two mechanisms to the
hysteresis. In this paper, we present electrometry measurements of a
bilayer 2D system obtained using a method recently developed by Ho
{\it et al.},~\cite{HoAPL10} and obtain a clear experimental
signature for the dominance of non-equilibrium induced current over
charge transfer (see Figs.~1(b/c) and 2). Our data clearly
demonstrates that charge transfer alone cannot be responsible for
the hysteresis effects previously reported in bilayer 2D
systems,~\cite{ZhuPRB00, TutucPRB03, PanPRB05, MisraPRB08} and is at
best a very small contribution towards this effect.

\begin{figure}
\includegraphics[width=8cm]{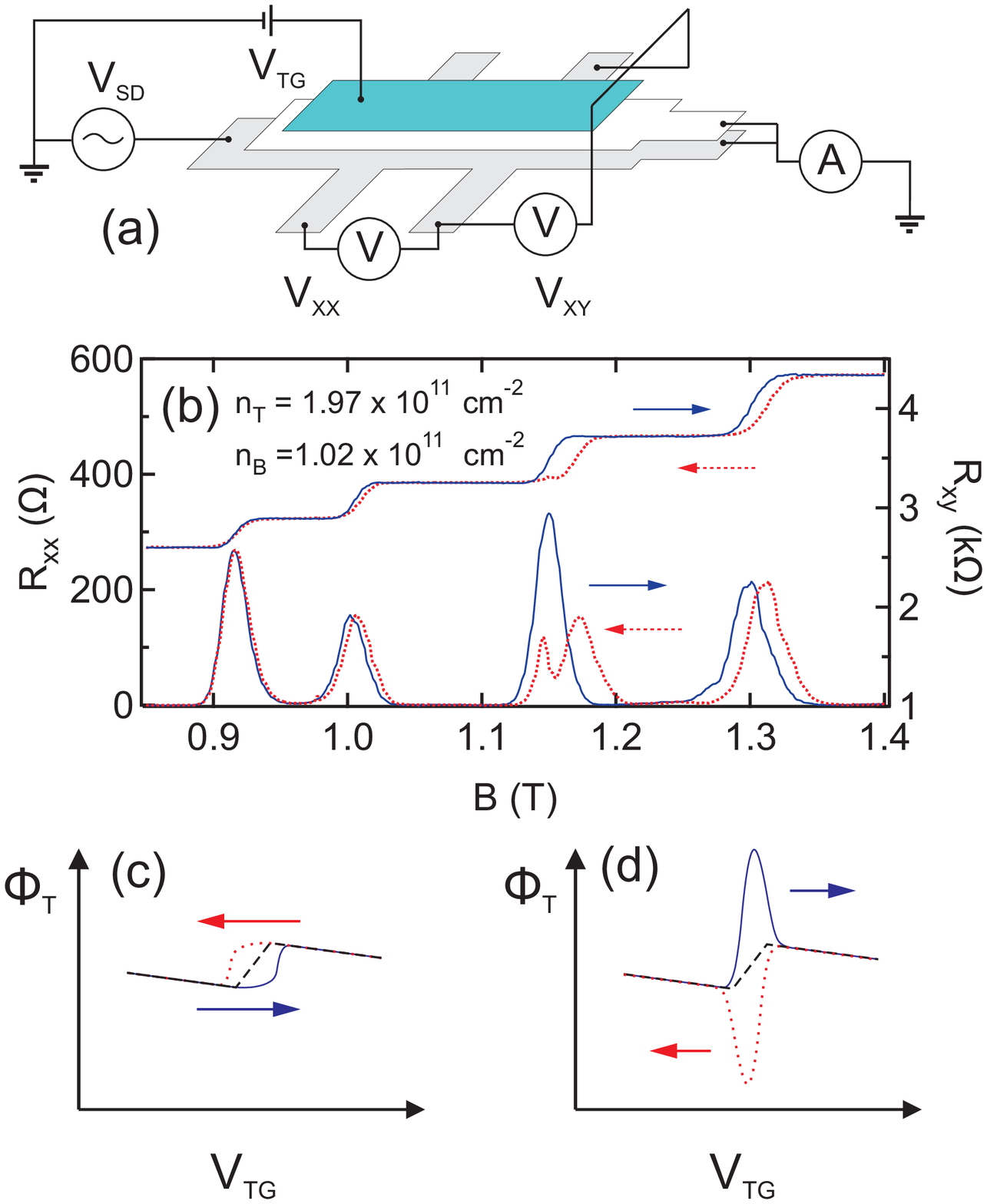}
\caption{\label{fig:schematic}} (color online) (a) A schematic of
the device and measurement circuit with (from bottom) the lower 2DES
in grey, the upper 2DES in white and the top-gate in blue. The upper
2DES shares only one common contact with the lower 2DES, which is
connected to ground, allowing the upper 2DES density to respond to
changes in top gate bias. The measured Hall voltage $V_{xy}$ can be
used to obtain the upper 2DES chemical potential $\mu_{T}$. (b) The
measured longitudinal resistance $R_{xx}$ (lower two traces) and
Hall resistance $R_{xy}$ (upper two traces) in the lower 2DES versus
magnetic field $B$ demonstrating similar hysteresis to that
previously reported in bilayer 2D systems.~\cite{ZhuPRB00, PanPRB05,
MisraPRB08} (c) and (d) are schematics illustrating the expected
behavior of the measured electrostatic potential $\phi_{T}$ in the
upper 2DES as a function of top-gate bias $V_{TG}$ for the charge
transfer and non-equilibrium induced current mechanisms
respectively. In (b)-(d) the blue solid/red dotted lines indicate
data taken with increasing/decreasing $B$ or $V_{TG}$.
\end{figure}

Our device was produced using a double quantum well heterostructure
featuring two $20$~nm wide GaAs quantum wells separated by a $30$~nm
Al$_{0.33}$Ga$_{0.67}$As barrier, giving an effective 2DES
separation of $d = 50$~nm. Standard semiconductor processing
techniques were used to produce a Hall-bar with NiGeAu ohmic
contacts that penetrate both quantum wells. The connections between
the upper 2DES and all but one of the ohmic contacts are severed
using a set of negatively biased `depletion' gates.~\cite{HoAPL10}
This isolates the upper 2DES from the measurement circuit, aside
from a connection to ground via the drain contact that enables the
upper 2DES density to be tuned by applying a voltage $V_{TG}$ to the
top-gate, as per the schematic in Fig.~1(a). With the top-gate
unbiased, the upper (lower) 2DES has a mobility of $1.2 \times
10^{6}$~cm$^{2}/$Vs ($1.4 \times 10^{6}$~cm$^{2}/$Vs) and density
$n_{\rm T} = 2.00 \times 10^{11}$~cm$^{-2}$ ($n_{\rm B} = 1.98
\times 10^{11}$~cm$^{-2}$). All electrical measurements were
performed at a temperature $\sim 50$~mK using four-terminal lock-in
techniques with an excitation voltage of $100~\mu$V at $17$~Hz.

We commence by demonstrating that our device shows hysteresis
similar to that previously observed in bilayer 2D
systems.~\cite{ZhuPRB00, TutucPRB03, PanPRB05, MisraPRB08} In
Fig.~1(b), we plot the longitudinal and Hall resistances $R_{xx}$
and $R_{xy}$ of the lower 2DES measured with increasing (blue solid
line) and decreasing (red dotted line) perpendicular magnetic field
$B$. To compare directly with previous work, we deliberately
imbalance the two 2DESs (i.e., $n_{T} \neq n_{B}$) when making this
measurement. We observe clear hysteresis in regions where $R_{xx}
\neq 0$, consistent with the literature, for example, cf. Fig.~1 of
Zhu {\it et al.},~\cite{ZhuPRB00} Fig.~2 of Pan {\it et
al.},~\cite{PanPRB05} or Fig.~4 of Misra {\it et
al}.~\cite{MisraPRB08}

We now turn to measurements of the electrostatic potential
$\phi_{T}$ in the upper 2DES. The lower `sensor' 2DES is used as a
capacitively-coupled electrometer,~\cite{HoAPL10} whereby changes in
$\phi_{T}$ lead to changes in the lower 2DES density $\Delta n_{B} =
\epsilon\Delta \phi_{\rm T} /(ed)$. The latter is observed by
monitoring $R_{xy}$ as the upper 2DES density $n_{\rm T}$ is varied
by sweeping $V_{TG}$ between $0$~V and the depletion of the upper
2DES at $V_{TG} = -0.3$~V. For maximum sensitivity, we choose a
fixed `operating point' in magnetic field $B$ where the lower 2DES
is at the midpoint between two Hall plateaus, and the relationship
between $R_{xy}$ and $n_{\rm B}$ is obtained by characterizing the
shape of this quantum Hall transition.~\cite{HoAPL10}

Changes in $\phi_{T}$ result from changes in the upper 2DES chemical
potential $\mu_{T}$, and the electrostatic potential $\phi_{NEC}$
due to non-equilibrium induced current flowing in the upper 2DES
such that $\Delta \phi_T = \Delta \mu_{T}/e + \phi_{NEC}$. Before
looking at the actual data, we first consider the expected behavior
of $\phi_T$ under the two possible mechanisms for the hysteresis at
fixed $B$. At complete equilibrium (i.e., there is no
non-equilibrium induced current $\phi_{NEC} = 0$ or charge transfer
hysteresis), $\mu_{T}$ would follow a sawtooth-shaped path (black
dashed line in Fig.~1(c/d)) with increasing $n_{T}$  or $V_{TG}$,
falling gradually at fixed Landau level occupation due to negative
compressibility, and rising rapidly at integer filling factor
$\nu_{T}$ due to repopulation of the lowest unoccupied Landau
level.~\cite{HoAPL10} In the charge transfer mechanism, the long
$RC$ time-constant should lead to a hysteretic $\mu_{T}$ resulting
in the behavior shown in the blue solid and red dashed lines in
Fig.~1(c). The most significant features are that $\mu_{T}$ evolves
monotonically between the potentials immediately before and after
the transition, and always lags the equilibrium transition point by
an amount directly proportional to the sweep rate. The expected
behavior for the non-equilibrium current mechanism is shown in
Fig.~1(d). Here $\phi_{T}$ leads the equilibrium transition point,
and significantly overshoots the potential on either side due to
$\phi_{NEC}$.~\cite{HuelsPRB04} In practice, this overshoot is
large, the hysteresis is an order of magnitude larger than the
equilibrium change in $\mu_{T}$.

\begin{figure}
\includegraphics[width=8cm]{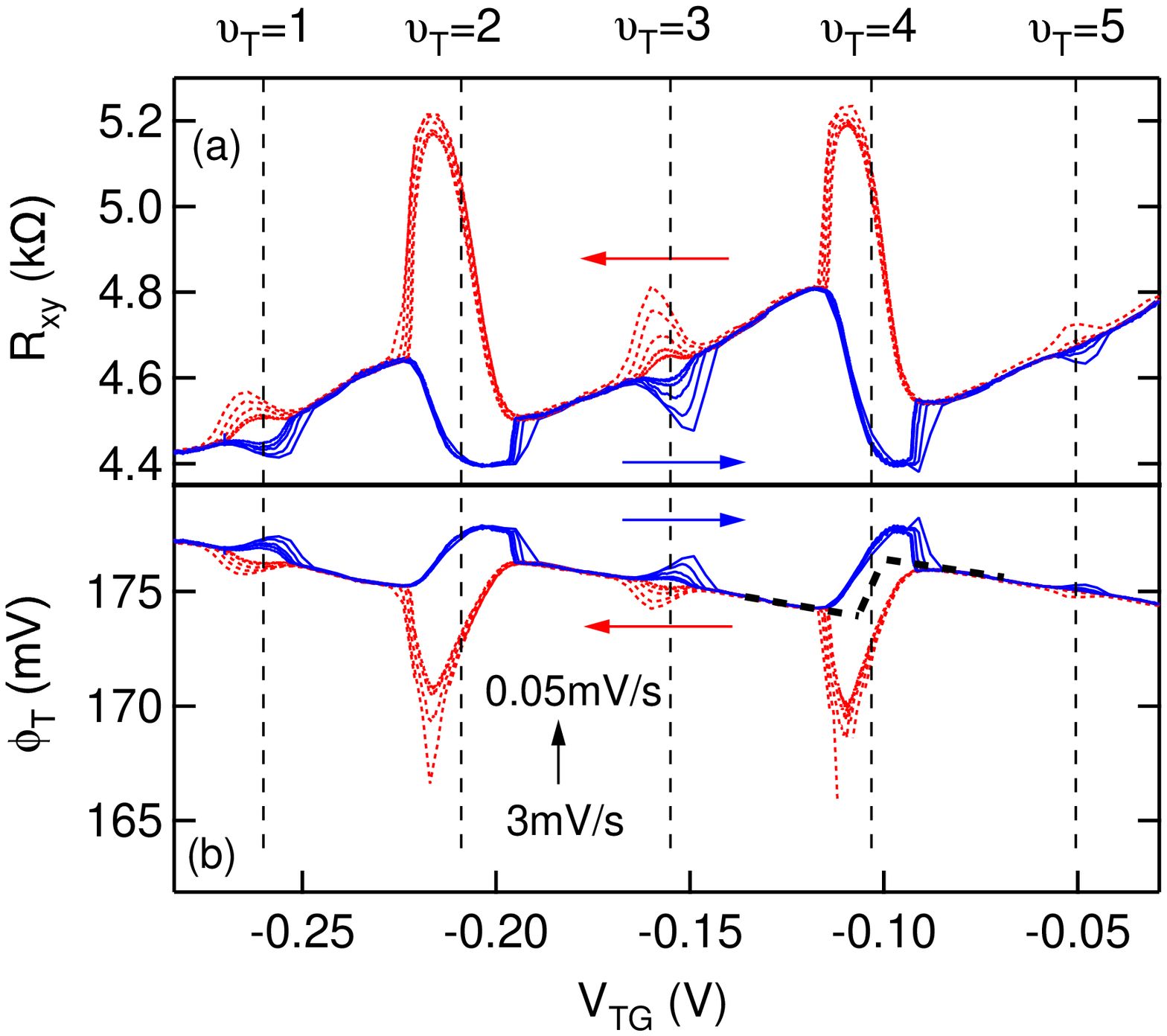}
\caption{\label{fig:sweeprate}} (color online) (a) The measured
lower 2DES Hall resistance $R_{xy}$ and (b) the corresponding upper
2DES electrostatic potential $\phi_{T}$ versus top-gate voltage $V_{TG}$
for decreasing $V_{TG}$ (red dotted lines) and increasing $V_{TG}$
(blue solid lines). The dashed black line in (b) highlights the
ideal behavior of $\mu_{T}$ in the absence of any hysteresis. In
each case, five traces are presented, obtained at $V_{TG}$ sweep
rates of $3$, $1$, $0.3$, $0.1$ and $0.05$~mV/s, with the largest
deviation from ideal behavior occurring at highest sweep-rate.
\end{figure}

In Fig.~2(a) and (b) we show the measured $R_{xy}$ and corresponding
$\phi_T$ versus $V_{TG}$ at a variety of sweep rates between $0.05$
and $3$~mV/s. The up-sweeps towards increased $V_{TG}$ are presented
with blue solid lines and the down-sweeps are presented with red
dashed lines. Hysteresis is clearly evident at integer filling
factor $\nu_{T}$ in Fig.~2, with the equilibrium change in chemical
potential highlighted by the dashed black line near $\nu_{T} = 4$ in
Fig.~2(b). The overshoot observed undeniably points towards
non-equilibrium induced current being generated in the upper
2DES.~\cite{FaulhaberPRB05}
This overshoot is also visible at odd $\nu_{T}$. Here the equilibrium changes in chemical potential are due to the Zeeman spin-splitting and are too small to be observed, however the hysteresis remains clearly visible.
Another notable feature is that the
hysteresis is significantly stronger at even $\nu_{T}$ than at odd
$\nu_{T}$. This is consistent with the non-equilibrium current
mechanism, since the larger change in $\mu_{T}$ that occurs at even
$\nu_{T}$ will produce more dissipationless edge states in the upper
2DES, which result in larger induced currents and correspondingly
stronger hysteresis. The same behavior is widely observed in
magnetometry~\cite{UsherJPCM09} and electrometry~\cite{HuelsPRB04}
measurements of single 2D systems.

To delve further into the data in Fig. 2, we must first consider how
the hysteresis should depend on rate of change of $V_{TG}$ in the
two mechanisms. To first order, charge transfer should give no sweep
rate dependence because the factor limiting charge migration from
one 2DES to the other is the very low conductivity of the bulk
states, which is independent of the rate of change in $V_{TG}$.
However, since the sweep rate also affects the rate at which the
equilibrium chemical potential in each 2DES changes, there will be a
weak sweep rate dependence in the charge transfer mechanism. In
contrast, the sweep rate dependence for the non-equilibrium current
mechanism should be strong. For the case where $B$ is swept rather
than $V_{TG}$, the explanation is quite simple -- the
non-equilibrium induced current in the edge states is just the eddy
current, as described by Faraday's law $\epsilon \propto -dB/dt$,
where $\epsilon$ is the induced electromotive force (emf). If we
sweep $V_{TG}$ instead, then the change in density results in charge
flowing into/out of the 2DES via the ohmic contacts. This flow is
initially directed towards the centre of the 2DES, but is rapidly
channeled into the edge states by the Lorentz force. We note that
the sweep rate dependence of real induced current will differ from
the simple pictures discussed above due to the breakdown of the
quantum Hall effect~\cite{UsherJPCM09} and also a capacitive
mechanism in gated samples.~\cite{RuhePRB09}

With the expected behavior in mind, at first sight, the measured
sweep rate dependence in Fig.~2(b) appears to be inconsistent with
the dominance of either mechanism -- the sweep rate dependence at
odd $\nu_{T}$ is strong but for even $\nu_{T}$ it appears to be very
weak. For example, in Fig.~2(a) the peak-peak amplitude of the
hysteresis `loop' at $\nu_{T} = 3$ varies from $53~\Omega$ at a
sweep rate $dV_{TG}/dt = 0.05$~mV/s to $334~\Omega$ at $dV_{TG}/dt =
3$~mV/s, an increase of over $500~\%$ at $\nu_{T} = 3$. In
comparison, for $\nu_{T} = 4$ the amplitude increases from
$786~\Omega$ to $857~\Omega$ for the same sweep rate change, the
corresponding difference being only $9~\%$. This lack of sweep rate
dependence at even $\nu_{T}$ is ultimately due to a limitation in
our electrometry technique. As mentioned earlier, we sense changes
in the electrostatic potential in the upper 2DES by setting a
magnetic field `operating point' such that the lower 2DES is
directly in between two Hall plateaus (i.e., to half integer filling
factor $\nu_{B}$ in the lower 2DES) and monitoring the lower 2DES
Hall resistance $R_{xy}$.~\cite{HoAPL10} While this gives us optimum
sensitivity, since $dR_{xy}/d\phi_{T}$ is a maximum at half integer
$\nu_{B}$, the range is limited because as $R_{xy}$ approaches the
adjacent Hall plateaus this sensitivity drops to zero. The data in
Fig.~2 was obtained at $B = 1.55$~T (i.e., $\nu_{B} = 5.5$), which
means that the lower 2DES `sensor' sensitivity is maximal at $R_{xy}
= 4693$~k$\Omega$ and diminishes monotonically to zero by $R_{xy} =
4302$ and $5162$~k$\Omega$. A careful look at the abscissa in
Fig.~2(a) reveals that the hysteresis loops at even $\nu_{T}$ have
reached these zero sensitivity limits, and thus the lack of
sweep-rate dependence in $R_{xy}$ does not imply a lack of sweep
rate dependence in the size of the induced current.

Overcoming this sensitivity problem is not straightforward. The most
obvious solution would be to shift the operating point to obtain a
reduced half-integer $\nu_{B}$ where the $R_{xy}$ range between the
adjacent Hall plateaus is greater. Unfortunately, the higher
magnetic field required to achieve this also reduces $\nu_{T}$,
pushing the data in Fig.~2(a) outside the available range of
$V_{TG}$. A better solution would be to introduce a back-gate in the
device so that the lower 2DES density can be tuned independently of
the upper 2DES density. Then it would be possible to implement a
feedback mechanism, whereby the lower 2DES is maintained at the
optimum half-integer filling factor with the magnetic field fixed,
with the back-gate voltage required to do used as the measurable
quantity in detecting changes in $\phi_{T}$. This will not be simple
to implement, and is the goal of future work; in this paper, we will
follow another avenue to demonstrate that there is in fact a strong
sweep rate dependence at even $\nu_{T}$ consistent with
non-equilibrium induced current being the dominant contribution to
the hysteresis.

\begin{figure}
\includegraphics[width=8cm]{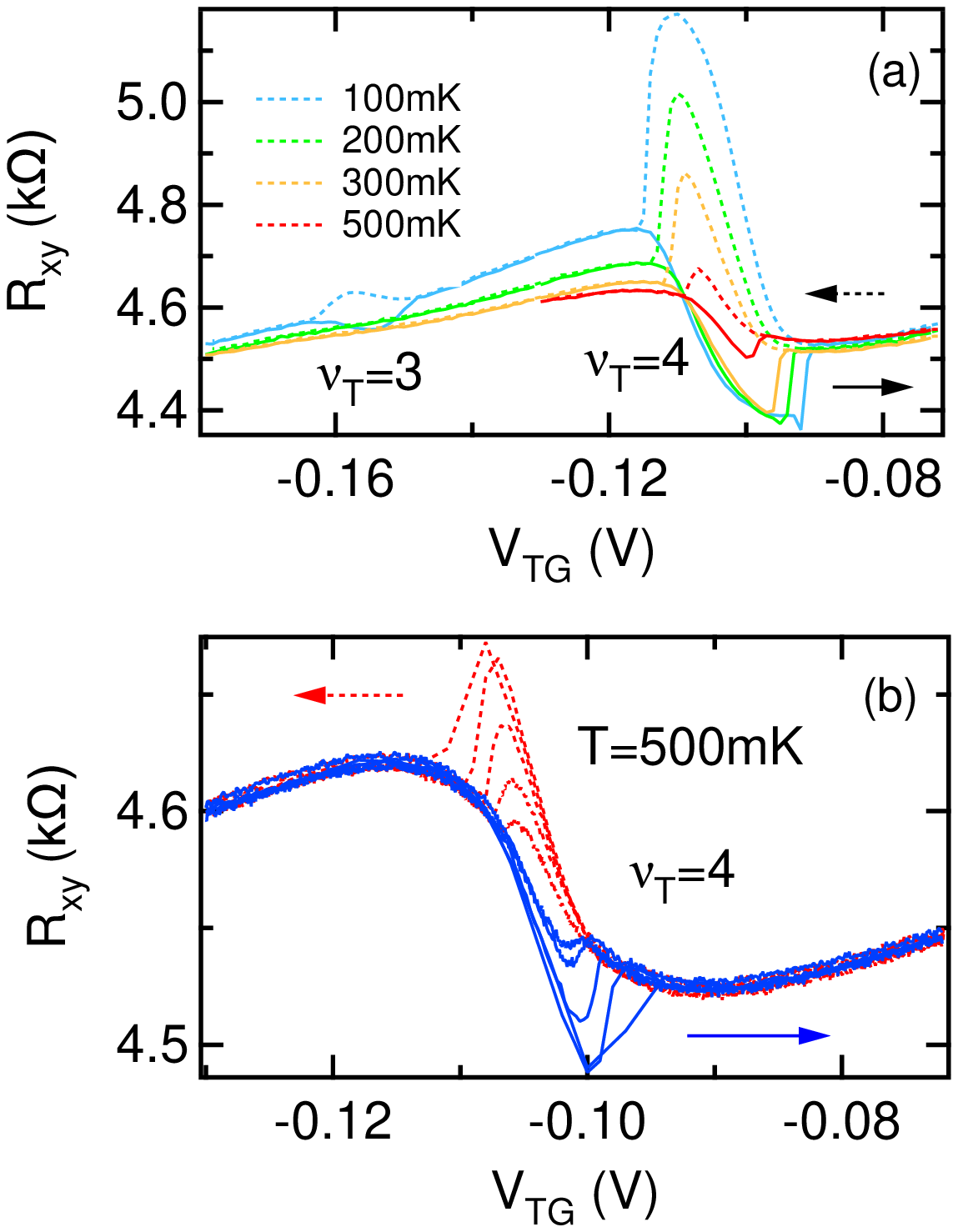}
\caption{\label{fig:tempdep}} The Hall resistance $R_{xy}$ vs
$V_{TG}$ obtained at $B = 1.55$~T for (a) a fixed sweep rate of
$1$~mV/s for temperatures $T$ of $100$, $200$, $300$ and $500$~mK,
and (b) a fixed temperature $T = 500$~mK for sweep rates of $0.05$,
$0.1$, $0.3$, $1$ and $3$~mV/s. In both panels,
up-sweeps/down-sweeps are indicated with solid/dashed lines.
\end{figure}

In Fig.~3(a), we present the temperature dependence of the
hysteresis loops at $\nu_{T} = 3$ and $4$ for a fixed sweep rate of
$1$~mV/s. As the temperature $T$ is increased towards $500$~mK, the
hysteresis loops at both odd and even $\nu_{T}$ diminish in
amplitude. This is also observed in magnetization measurements of a
single 2DES in the quantum Hall regime,~\cite{RuhePRB09} and is due
to increasing dissipation in the edge states with
temperature.~\cite{TsuiPRB82,CagePRB84} However, this also carries a
positive side-effect, which is a small increase in the dynamic range
of the lower 2DES as a sensor for the upper 2DES electrostatic
potential. The hysteresis at $\nu_{T} = 3$ is quenched by $T =
200$~mK, and while the hysteresis at $\nu_{T} = 4$ remains at $T =
500$~mK, it is sufficiently diminished that it stays well within the
sensitivity range of the upper 2DES. Thus we repeat the sweep rate
dependence study for $\nu_{T} = 4$ at $T = 500$~mK, the results are
presented in Fig.~3(b). Here the strong sweep rate dependence
expected for the non-equilibrium induced current is indeed observed,
with the peak-to-peak amplitude increasing by $243~\%$ as the sweep
rate is increased from $0.05$ to $3$~mV/s. This confirms that the
lack of sweep rate dependence for even $\nu_{T}$ in Fig.~2 was due
to the large induced current driving the sensor 2DES into
insensitive regions.

In conclusion, we have studied the origin of the hysteresis observed
in bilayer 2D systems in the quantum Hall regime using a newly
developed electrometry technique~\cite{HoAPL10} that allows direct
measurements of the electrostatic potential in one of the two 2D
systems. We observe hysteresis `loops' that point very strongly to
the dominance of non-equilibrium induced current over impeded charge
transfer as a cause of the hysteresis recently reported in bilayer
2D systems.~\cite{ZhuPRB00, TutucPRB03, PanPRB05, MisraPRB08} Our
findings are consistent with previous measurements of hysteresis in
single 2D systems.~\cite{UsherJPCM09,HuelsPRB04}

This work was funded by Australian Research Council (ARC).  L.H.H.
acknowledges financial support from the UNSW and the CSIRO. ARH and
APM acknowledge financial support from ARC Professorial and Future
Fellowships, respectively. The authors thank Dr Jack Cochrane for
technical support. This work was performed in part using the NSW
node of the Australian National Fabrication Facility (ANFF).


\begin{thebibliography}:

\bibitem{vonKlitzingPRL80} K.v. Klitzing, G. Dorda and M. Pepper, Phys. Rev. Lett. {\bf 45}, 494 (1980).

\bibitem{TsuiPRL82} D.C. Tsui, H.L. St\"{o}rmer and A.C. Gossard, Phys. Rev. Lett. {\bf 48}, 1559 (1982).

\bibitem{EisensteinNat04} J.P. Eisenstein and A.H. MacDonald, Nature {\bf 432}, 691 (2004).

\bibitem{GirvinPhysTod00} S.M. Girvin, Physics Today {\bf 53(6)}, 39 (2000).

\bibitem{PiazzaNat99} V. Piazza, V. Pellegrini, F. Beltram, W. Wegscheider, T. Jungwirth and A.H. MacDonald, Nature {\bf 402}, 638 (1999).

\bibitem{dePoortereSci00} E.P. De Poortere, E. Tutuc, S.J. Papadakis and M. Shayegan, Science {\bf 290}, 1546 (2000).

\bibitem{ZhuPRB00} J. Zhu, H.L. Stormer, L.N. Pfeiffer, K.W. Baldwin, and K.W. West, Phys. Rev. B {\bf 61}, R13361 (2000).

\bibitem{TutucPRB03} E. Tutuc, R. Pillarisetty, S. Melinte, E.P. De Poortere, and M. Shayegan, Phys. Rev. B {\bf 68}, 201308(R) (2003).

\bibitem{PanPRB05} W. Pan, J.L. Reno, and J.A. Simmons, Phys. Rev. B {\bf 71}, 153307 (2005).

\bibitem{MisraPRB08} S. Misra, N.C. Bishop, E. Tutuc, and M. Shayegan, Phys. Rev. B {\bf 78}, 035322 (2008).

\bibitem{UsherJPCM09}  A. Usher and M. Elliott, J. Phys.: Condens. Matter {\bf 21}, 103202 (2009).

\bibitem{WattsPRL98} J.P. Watts, A. Usher, A.J. Matthews, M. Zhu, M. Elliott, W.G. Herrenden-Harker, P.R. Morris, M.Y. Simmons, and D.A. Ritchie, Phys. Rev. Lett. {\bf 81}, 4220 (1998).

\bibitem{FaulhaberPRB05} D.R. Faulhaber and H.W. Jiang, Phys. Rev. B {\bf 72}, 233308 (2005).

\bibitem{HuelsPRB04} J. Huels, J. Weis, J. Smet, K.v. Klitzing and Z.R. Wasilewski, Phys. Rev. B {\bf 69}, 085319 (2004).

\bibitem{HoAPL10} L.H. Ho, L.J. Taskinen, A.P. Micolich, A.R. Hamilton, P. Atkinson, M. Pepper and D.A. Ritchie, Appl. Phys. Lett. {\bf 96}, 212102 (2010).


\bibitem{RuhePRB09} N. Ruhe, G. Stracke, Ch. Heyn, D. Heitmann, H. Hardtdegen, Th. Sch\"{a}pers, B. Rupprecht, M.A. Wilde and D. Grundler, Phys. Rev. B {\bf 80}, 115336 (2009).

\bibitem{TsuiPRB82} D.C. Tsui, H.L. St\"{o}rmer and A.C. Gossard, Phys. Rev. B {\bf 25}, 1405 (1982).

\bibitem{CagePRB84} M.E. Cage, B.F. Field, R.F. Dziuba, S.M. Girvin, A.C. Gossard and D.C. Tsui, Phys. Rev. B {\bf 30}, 2286 (1984).












\end{thebibliography}
\end{document}